\documentstyle[12pt]{article}
\def\x{\stackrel{\otimes}{,}}

\def\a{\begin{eqnarray}}
\def\b{\end{eqnarray}}
\def\0{\nonumber}
\input amssym.def
\font\teneusm=eusm10                    % loads Euler script medium weight
\font\seveneusm=eusm7                   % (Euler frak (eufm) already loaded
\font\fiveeusm=eusm5                    % by amssym.def)
\newfam\eusmfam
\textfont\eusmfam=\teneusm
\scriptfont\eusmfam=\seveneusm
\scriptscriptfont\eusmfam=\fiveeusm

\input amssym
\newcommand{\be}{\begin{equation}}
\newcommand{\ee}{\end{equation}}
\newcommand{\ba}{\begin{eqnarray}}
\newcommand{\ea}{\end{eqnarray}}
\newcommand{\ban}{\begin{eqnarray*}}
\newcommand{\ean}{\end{eqnarray*}}
\newcommand{\brr}{\begin{array}}
\newcommand{\err}{\end{array}}
\newcommand{\bc}{\begin{center}}
\newcommand{\ec}{\end{center}}
\newcommand{\sss}{\scriptscriptstyle}

\def\A{{\cal A}}
\def\B{{\cal B}}
\def\D{{\cal D}}
\def\E{{\cal E}}
\def\G{{\cal G}}
\def\H{{\cal H}}

\def\N{{\cal N}}
\def\S{{\cal S}}
\def\P{{\cal P}}
\def\Q{{\cal Q}}
\def\T{{\cal T}}
\def\V{{\cal V}}
\def\W{{\cal W}}
\def\l{\lambda}
\def\s{\sigma}
\def\al{\alpha}
\def\o{\otimes}
\newcommand{\bea}{\begin{eqnarray}}
\newcommand{\eea}{\end{eqnarray}}
\newcommand{\bean}{\begin{eqnarray*}}
\newcommand{\eean}{\end{eqnarray*}}
\newcommand{\Slc}{SL_2({\Bbb C}\, )}
\newcommand{\Slr}{SL_2({\Bbb R}\, )}

\newcommand{\Slnr}{SL_n({\Bbb R}\, )}

\newcommand{\dpiu}{\partial_+}
\newcommand{\dmeno}{\partial_-}

\newcommand{\rref}[1]{(\ref{#1})}
\begin{document}
\begin{flushright}
SISSA-ISAS 210/92/EP
\end{flushright}
\vskip0.5cm
\centerline{\LARGE Free field representation of Toda field theories}
\vskip1.5cm
\centerline{\large  E. Aldrovandi\footnote{Address after November 1, 1992:}}
\centerline{International School for Advanced Studies (SISSA/ISAS)}
\centerline{Via Beirut 2, 34014 Trieste, Italy}
\centerline{\large  L. Bonora, V.Bonservizi, R.Paunov}
\centerline{International School for Advanced Studies (SISSA/ISAS)}
\centerline{Via Beirut 2, 34014 Trieste, Italy}
\centerline{and INFN, Sezione di Trieste.  }
\vskip5cm
\abstract{We study the following problem: can a  classical
$sl_n$ Toda field theory
be represented by means of free bosonic oscillators through a Drinfeld--Sokolov
construction? We answer affirmatively in the case of a cylindrical space--time
and for real hyperbolic solutions of the Toda field equations. We establish
in fact a one--to--one correspondence between such solutions and the space
of free left and right bosonic oscillators with coincident zero modes. We
discuss the same problem for real singular solutions with non hyperbolic
monodromy.}
\vfill
\eject

\section{Introduction}

Toda field theories in 2D have become a favorite research topic in theoretical
physics. One reason for this interest lies in the fact that Toda field theories
 (ToFT's) based on classical finite dimensional Lie algebras (in this paper
we will be dealing only with the latter) underlie many remarkable
conformal field theory models -- in particular all sort of minimal models --;
in fact they possess both the necessary conformal structure, which manifests
itself for example in the chiral splitting, and the integrable structure,
which is connected with the hidden quantum group symmetry typical of
these models. Both structures are simultaneously expressed in an elegant form
by the exchange algebra.

Another reason of interest is the connection of the $sl_2$ ToFT,
i.e. the Liouville theory (\cite{GN},\cite{CT},\cite{S},
with string theory and 2D gravity, together
with the evoked possibility that $sl_n$ ToFT might bear a relation
to 2D gravity coupled to conformal matter in much the same way as the
latter combination appears in matrix  models. The present state of affairs
does not even allow us to exclude that there might be a direct connection
between matrix models and Liouville or Toda theories formulated on the lattice.

Toda field theories can be regarded as Hamiltonian reductions of WZNW
models -- this is in itself a vast field of research -- or -- as we do in this
paper -- as autonomous field theories characterized by a W--algebra symmetry.
They are in fact at the origin of the present
interest in W--algebras. The geometrical meaning of
W--algebras is still rather obscure. However if we think that
the Liouville equation is well--known
to be the basis of the uniformization theory of Riemann surfaces, it is
not unmotivated to expect geometry to play a deep role in
ToFTs and, viceversa, that the latter might lead to significant
geometrical developments.

All this sounds pretty appealing to all those who have followed the most recent
developments in theoretical physics. On the other hand, even though the
research
in this field has been intensive (\cite{GB},\cite{FL},\cite{B},\cite{GM}),
many questions in Toda
field theories are still unanswered. Among the latter we quote
in particular the problem of constructing conformal blocks in W--algebra.
We feel a more thorough analysis of ToFT's is necessary.
This motivated us in taking up this
research, which could be synthesized as the search for an answer to the
following {\it question: to what extent can we represent ToFTs
by means of free bosonic oscillators ?}  In this sense this paper is the
continuation of \cite{BBT}. There it was shown that to every ToFT
we can associate two (one for each chirality) Drinfeld--Sokolov (DS) linear
systems defined in terms of independent free bosonic oscillator fields. By
means
of these we can construct solutions of the ToFT equations of motion. The
question left unanswered in \cite{BBT} was: Do we construct in this
way all the periodic and local solutions of the ToFT equations of motion
so that the Poisson
brackets for the bosonic oscillators correspond to the canonical Poisson
brackets in the ToFT?

In this paper we address the above two problems for an $sl_n$ ToFT defined in a
cylindrical space--time, where space is represented by a circle and time by a
straight line, therefore with periodic space boundary conditions.
The most important result is the following: we construct
a one--to--one correspondence between {\it real hyperbolic} solutions\footnote{
Hyperbolic solution means a solution with hyperbolic monodromy. All regular
enough solutions are hyperbolic. For a more precise statement
see Appendix A and B.}
of  ToFT equations and
appropriate set of free bosonic oscillator fields of the DS linear systems
which preserve the canonical symplectic structure of the ToFT.
We realize in this way a parametrization of the classical phase
space of the ToFT's, which lends itself to canonical quantization.
In conclusion the above question has the following {\it answer:
ToFT's can be fully represented by means of free bosonic oscillators},
with the limitations expressed above.

On the one hand, this conclusion can be considered, as a
completion of a program started by Leznov and Saveliev \cite{LS} and
continued in \cite{BBT}.
On the other hand the same problem is still open
for more complicated topologies (see \cite{AB} for a discussion
over Riemann surfaces).

We partially analyze also the (singular) solutions of parabolic
monodromy. For a large family of them it is also possible to define a free
field representation. The role of these solutions in quantizing the theory
is unclear.

The paper is organized as follows.
Since we wish the paper to be as self--contained as possible, in section 2
we review ref.\cite{BBT} and state the problem announced above in a more
precise form. In section 3 we show how to find the solution
for the Liouville theory. In section 4 we extend the proof to an
$sl_n$ Toda field theory. Section 3 is used as a general guide and
presented in a very detailed form, while section 4 is more sketchy.
In section 5 we discuss the family of singular solutions of the Liouville
equation referred to above.

\section{The problem}

Let ${\cal G}$ be a simple finite dimensional Lie algebra of rank $r$,
equipped with an invariant scalar product denoted by $(~,~)$. We choose a
Cartan subalgebra  ${\cal H}$ with an orthonormal basis $\{ H_i \}$.
We recall the following commutation relations in a Cartan-Weyl basis
\begin{eqnarray}
\relax [H,E_{\pm \alpha}]&=& \pm \alpha (H) E_{\pm \alpha}
\nonumber \\
\relax [E_{\alpha},E_{-\alpha}] &=& H_\alpha
\nonumber
\end{eqnarray}
Toda field theories are defined by means of a linear system.
\begin{eqnarray}
(\partial_{\pm} +A_{\pm })T=0
\label{LS}
\end{eqnarray}
where $x_\pm=x\pm t$ are the light cone coordinates, and $\partial_\pm\equiv
\partial_{x_\pm}=
\textstyle{1 \over 2}(\partial_x \pm \partial_t)$.
\begin{eqnarray}
A_{+}=\partial_{+}\Phi + e^{ad \Phi}{\cal E}_+, \quad\quad\quad
A_{-}=-\partial_{-}\Phi + e^{-ad \Phi}{\cal E}_-
\0
\end{eqnarray}
The field $\Phi$ takes values in the Cartan subalgebra  and
\begin{eqnarray}
{\cal E}_+ &=& \sum_{\alpha ~simple} E_{\alpha} \nonumber \\
{\cal E}_- &=& \sum_{\alpha ~simple} E_{-\alpha} \nonumber
\end{eqnarray}
The zero curvature condition
\begin{eqnarray}
F_{+-}=\partial_{+}A_{-}-\partial_{-}A_{+}+[A_{+},A_{-}]
=0 \nonumber
\end{eqnarray}
yields the equations of motion
\begin{eqnarray}
\partial_{+}\partial_{-} \Phi =
{\textstyle {1\over 2}}\sum_{\alpha \;\;simple} e^{2 \alpha (\Phi)} H_\alpha
\label{Toda}
\end{eqnarray}

There is a standard way to represent solutions of eq.(\ref{Toda}).
Given a highest weight vector $|\Lambda^{(r)}>$, we define
\begin{eqnarray}
\xi^{(r)}(x)&=&<\Lambda^{(r)}|e^{-\Phi(x)}T(x) \nonumber \\
\overline{\xi}^{(r)} (x)&=& T^{-1}(x)e^{-\Phi(x)}|\Lambda^{(r)}>
\label{xi}
\end{eqnarray}
where $T(x)$ is the transport matrix
\begin{eqnarray}
T(x)=P\exp\left( -\int_0^x A_x dx\right)
\nonumber
\end{eqnarray}
and $A_x=A_{+}+A_{-}$.
$T(x)$ is the solution of the equation
$ (\partial_x+A_x)T=0$
with the initial condition $T(0)=1$.

Using the explicit form of $A_{+}$ and $A_{-}$ one can easily show that
$\xi$ and ${\overline \xi}$ are chiral objects, i.e.
\begin{eqnarray}
\partial_{-}\xi=0&\quad&\qquad \partial_{+}{\overline \xi}=0\nonumber
\end{eqnarray}
The objects defined by eqs.(\ref{xi}) satisfy the exchange algebra
\begin{eqnarray}
\{\xi^{(r)}(x) \x \xi^{(r')}(y)\}&
=&\xi^{(r)}(x)\otimes\xi^{(r')}(y)[\theta(x-y)r^+ +\theta(y-x)r^-]\nonumber \\
\{\overline{\xi}^{(r)}(x)\x\overline{\xi}^{(r')}(y)\}&
=&
[\theta(x-y)r^-+\theta(y-x)r^+]\overline{\xi}^{(r)}(x)
\otimes\overline{\xi}^{(r')}(y) \nonumber \\
\{\xi^{(r)}(x)\x\overline{\xi}^{(r')}(y)\}&
=&-\xi^{(r)}(x)\otimes 1\cdot r^- \cdot 1\otimes
\overline{\xi}^{(r')}(y)\nonumber \\
\{\overline{\xi}^{(r)}(x)\x\xi^{(r')}(y)\}&
=&-1 \otimes \xi^{(r')}(y)\cdot  r^+ \cdot
\overline{\xi}^{(r)}(x)\otimes 1 \label{exchange}
\end{eqnarray}
where $r^{\pm}$ are the solutions of the classical Yang-Baxter equation
\a
&&r^+=t_0 +2 \sum _{\alpha~positive} {{E_\alpha \otimes E_{-\alpha}}\over
{(E_\alpha, E_{-\alpha})}}\label{r+}\\
&&r^-=- t_0-2 \sum _{\alpha~positive}
{{E_{-\alpha} \otimes E_{\alpha}}\over
{(E_{-\alpha}, E_{\alpha})}}\label{r-}\0
\b
and
\a
t_0=\sum_i H_i\otimes H_i\0
\b

This is a consequence of the canonical Poisson bracket defined by
\a
\{\pi_\Phi(x)\x \Phi(y) \} = \delta (x-y)\cdot t_0 \label{CPB}
\b
$\pi_\Phi$ being the canonical momentum of $\Phi$, i.e. $\pi_\Phi =
\partial_t \Phi$.

If the Toda field $\Phi(x)$ is periodic, the $\xi$, $\overline{\xi}$ fields
have the monodromy properties
\begin{eqnarray}
\xi^{(r)}(x+2\pi ) &=& \xi^{(r)}(x) ~T \nonumber \\
\overline{\xi}^{(r)}(x+2\pi ) &=& T^{-1} ~\overline{\xi}^{(r)}(x)
\0
\end{eqnarray}
where $T=T(2\pi)$. We also recall the following Poisson brackets
\begin{eqnarray}
\{T \x T \} &=& - [r_{12}^\pm , T\otimes T ]  \nonumber \\
\{\xi^{(r)}(x) \x T \} &=&\xi^{(r)}(x) \otimes T \cdot r^- \label{Txi}\\
\{\overline{\xi}^{(r)}(x) \x T \} &=& - 1\otimes T \cdot r^+ \cdot
\overline{\xi}^{(r)}(x) \otimes 1
\0
\end{eqnarray}
Finally, from eq.(\ref{xi}), we see that the fields of Toda theory,
are reconstructed by means of the formula
\begin{eqnarray}
e^{-2\lambda^{(r)}(\Phi
)(x_+,x_-)}=\xi^{(r)}(x_+)\cdot\overline{\xi}^{(r)}(x_-)
\label{recon}
\end{eqnarray}
where $\lambda^{(r)}$ is the weight corresponding to $|\Lambda^{(r)}>$.
These fields are obviously periodic and  one can check that they are local,
i.e. they Poisson commute at equal time.

Presented in this form all the matter sounds a bit tautological, since
we have to already know the solutions of the Toda field equations in order
for the above formulas (in particular eq.(\ref{recon})) to work. In this
way we do not get any explicit representation of the space of
classical solutions. To achieve this we have to proceed in a different way,
i.e. consider separately the two chiral halves of the theory
and introduce the so--called associated Drinfeld--Sokolov  linear systems
\cite{DS}:
\begin{eqnarray}
\partial_{+} {\cal Q}_+- (P-{\cal E}_+ ){\cal Q}_+ =0, \quad\quad\quad
\partial_{-} {\cal Q}_+=0\label{DS+}
\b
\a
\partial_{+} {\cal Q}_-=0,\quad\quad\quad
\partial_{-} {\cal Q}_- +{\cal Q}_- (\overline{P}-{\cal E}_- )=0\label{DS-}
\end{eqnarray}
where ${\cal Q}_+(x_+)$ and ${\cal Q}_-(x_-)$
takes values in a Lie group
whose Lie algebra is ${\cal G}$, and
where $P$ and $\overline{P}$ are periodic chiral and antichiral, respectively,
fields which take
values in the Cartan subalgebra and have the Poisson brackets
\begin{eqnarray}
\{P(x) \x P(y) \}& =&- (\partial_x - \partial_y ) \delta(x-y) t_0\0\\
\{P(x) \x \overline{P}(y) \} &=&0 \label{PP} \\
\{\overline{P}(x) \x \overline{P}(y) \} &=& (\partial_x - \partial_y )
\delta(x-y) t_0
\0
\end{eqnarray}
At times we will refer to $P$ and $\bar P$ as DS fields.
{}From the solution ${\cal Q}_+(x)$ and ${\cal Q}_-(x)$
of eqs.(\ref{DS+},\ref{DS-})
normalised by ${\cal Q}_+(0)=1$, ${\cal Q}_-(0)=1$ we define a basis
$\sigma ,~\overline{\sigma}$
\begin{eqnarray}
\sigma^{(r)}(x) &=& <\Lambda^{(r)}|{\cal Q}_+(x) \label{sigmar} \\
\overline{\sigma}^{(r)}(x) &=& {\cal Q}_-(x) |\Lambda^{(r)}> \label{sigmabarr}
\end{eqnarray}
This basis has the Poisson bracket algebra
\begin{eqnarray}
\{\sigma^{(r)}(x) \x \sigma^{(r')}(y)\}&
=&\sigma^{(r)}(x)\otimes\sigma^{(r')}(y)[\theta(x-y)r^+ +\theta(y-x)r^-]
\label{sigmasigma}\\
\{\overline{\sigma}^{(r)}(x)\x\overline{\sigma}^{(r')}(y)\}&
=&
[\theta(x-y)r^-+\theta(y-x)r^+]\overline{\sigma}^{(r)}(x)\otimes
\overline{\sigma}^{(r')}(y) \label{sigmabarsigmabar}
\end{eqnarray}
while
\begin{eqnarray}
\{\sigma^{(r)} (x) \x \overline{\sigma}^{(r')} (y) \} =0 \label{sigmasigmabar}
\end{eqnarray}
Since $P(x)$ and $\bar P(x)$ are periodic we can expand them in Fourier
series
\a
P(x) = \sum_n P_n e^{inx}, \quad\quad \bar P(x) =\sum_n \bar P_n e^{inx}\0
\b
An important role in the following is played by
the left and  right monodromy matrices
\begin{eqnarray}
S={\cal Q}_+(2\pi) ,~~~~~~~~~~\bar S= {\cal Q}_-(2\pi) \nonumber
\end{eqnarray}
We will also need $K$ and $\bar K$, defined as follows
\begin{eqnarray}
K= \sum_{n\neq 0} {iP_n \over n}, \quad\quad\quad
\bar K= \sum_{n\neq 0} {i\bar P_n \over n} \nonumber
\end{eqnarray}

The aim now is to construct periodic local solutions of the the Toda field
equations (\ref{Toda}), exactly as the formal solutions (\ref{recon}) are,
but in terms of the free bosonic fields $P$ and $\bar P$.
A solution to eqs.(\ref{Toda}) is given by
\begin{eqnarray}
e^{-2\lambda^{(r)}(\Phi )(x_+,x_-)}= \sigma^{(r)}(x_+) M \bar\sigma^{(r)}(x_-)
\label{sigmasol}
\end{eqnarray}
where $M$ is a constant matrix to be determined.
Since
\begin{eqnarray}
\sigma^{(r)}(x+\textstyle 2\pi) = \sigma^{(r)}(x) S,~~~~~~~~~~~
\bar\sigma^{(r)}(x+\textstyle 2\pi) = \bar S ~\bar\sigma^{(r)}(x)\nonumber
\end{eqnarray}
to get a periodic solution, we must have
\begin{eqnarray}
S~ M~ \overline{S} =M \label{SMSbar}
\end{eqnarray}
In order to satisfy this equation one proceeds to
diagonalize the monodromy matrices
\begin{eqnarray}
&&S= g_S\kappa g_S^{-1},\quad \quad \kappa = e^{2\pi P_0}\label{diag}\\
&&\bar S = \bar g_S \bar \kappa \bar g_S^{-1}, \quad
\quad \bar \kappa = e^{-2\pi\bar P_0}\nonumber
\end{eqnarray}
Then condition (\ref{SMSbar}) will be satisfied if
\begin{eqnarray}
&&M=g_S {\cal D} {\bar g_S^{-1}} \nonumber \\
&&\kappa \bar \kappa = 1 \nonumber
\end{eqnarray}
where $ {\cal D}\in \exp({\cal H}) $, $~{\cal H}$ being
the Cartan subalgebra.
The second condition simply means that
$P_0=\overline{P}_0$ and will eventually be imposed. The diagonal
matrix ${\cal D}$ has to be chosen in such a way that the fields
of the LHS of eq.(\ref{SMSbar}) be local. The solution is
\begin{eqnarray}
{\cal D}= { \Theta} \overline{ \Theta},~~~~{\Theta} = e^{Q- K},~~~~
\overline{ \Theta} = e^{\bar Q +\bar K} \nonumber
\end{eqnarray}
and $Q, \bar Q$ are the conjugate variables  of $P_0$ and $\bar P_0$,
respectively. Therefore
\begin{eqnarray}
&&\{Q \x P_0 \}= \textstyle {1\over {\pi}} t_0 \label{defQ} \\
&&\{\overline{Q}  \x  \overline{P}_0 \}= \textstyle {1\over
{\pi}} t_0\nonumber
\end{eqnarray}
Finally we can write
\begin{eqnarray}
e^{-2\lambda^{(r)}(\Phi)(x_+,x_-)} =
\psi^{(r)}(x_+)~ \overline{\psi}^{(r)}(x_-) \label{solution}
\end{eqnarray}
where we define the new objects (Bloch wave basis)
\begin{eqnarray}
\psi^{(r)}(x) =\sigma^{(r)}(x) g_S { \Theta}, \quad\quad \quad \bar
\psi^{(r)}(x) =\bar {\Theta} \bar g_S ^{-1}\bar\sigma^{(r)}(x)
\label{defpsi}
\end{eqnarray}

The $\psi$ and $\bar \psi$  have diagonal monodromy $\kappa$ and $\bar \kappa$,
respectively, and obey the exchange algebra
\begin{eqnarray}
\{\psi^{(r)}(x) \x \psi^{(r')}(y)\}\, &=&\, -\textstyle {1\over 2}
\psi^{(r)}(x) \otimes
\psi^{(r')}(y)
\Big\{\epsilon (x-y) (r^+ -r^- ) - \label{psipsi}\\
&&\, -coth( \pi ad_1 P_0)
(r^+ -t_0) - coth (\pi ad_2P_0 )(r^- + t_0)\Big\}
\nonumber \\
\{\overline{\psi}^{(r)}(x) \x \overline{\psi}^{(r')}(y)\}\, &=&\, \textstyle
{1\over 2}
\Big\{\epsilon (x-y) (r^+ -r^- )  +coth( \pi ad_1 \overline{P}_0)
(r^- +t_0) \label{psibarpsibar}\\
&&\, +coth (\pi ad_2\overline{P}_0 )(r^+ - t_0)\Big\}
\overline{\psi}^{(r)}(x) \otimes \overline{\psi}^{(r')}(y) \nonumber
\end{eqnarray}
and, as long as the $P_0$ and $\bar P_0$ are considered independent, we also
have
\begin{eqnarray}
\{ \psi ^{(r)}(x) \x \overline{\psi}^{(r')}(y) \}=0 \label{psipsibar}
\end{eqnarray}

It is now easy to prove that eq.(\ref{solution}) represents a general
solution of eqs.(\ref{Toda}) which is both periodic and local
provided we reduce the phase space by imposing $P_0= \bar P_0$.

It is clear that the free bosonic oscillators $P_n$ and $\bar P_n$
(together with $Q$ and $\bar Q$) provide
a parametrization of the space of classical periodic and local solutions
of the Toda field equations. One can also verify by means of
(\ref{solution}) that (\ref{PP}) implies (\ref{CPB}).
However there still remain an unanswered question:

{\it Does the above construction ({\rm referred to in the following as the
DS construction}) exhaust all the periodic and local solutions
of the Toda field equations,
so that the canonical Poisson brackets (\ref{CPB}) match the Poisson
brackets (\ref{PP}) for the free bosonic oscillators?}

The rest of the paper is devoted to answering this question.
Before we end this section two remarks are in order:

{\bf Remark 1}. The DS construction could also proceed in a slightly different
manner. We could first identify $P_0$ with $\bar P_0$ and then introduce
${\cal D}=  e^{Q+\bar K - K}$ in $\sigma^{(r)}g_S {\cal D}
\bar g_S^{-1} \bar \sigma^{(r)}$
together with $Q=\bar Q$ in eq.(\ref{defQ}). All the above conclusions
would hold except, of course, eq.(\ref{psipsibar}). This remark will be useful
in the next section. In fact, since in the next section we reconstruct the
DS system starting from a periodic solution $\varphi$, we expect
the left and right zero modes to coincide.

{\bf Remark 2}. The problem studied in the next sections is not completely
new in the literature. At least for the Liouville model two fields $p$ and
$\tilde p$ were found in \cite{GN} that correspond to the two solutions
for the $p$ field found in the next section. However that
result is limited to the open string case, which has particular features.

\subsection{Description of the method for solving the problem}

The method for solving the problem consists of two steps.

The most natural (but, as it will turn out, insufficient) idea is to
perform successive
field--dependent gauge transformations on the initial linear system (\ref{LS})
so as to reduce it to the DS form -- we will see later on that this corresponds
to the Gauss decomposition of the transport matrix \cite{LS}.
Let us consider first the following transformations
\a
V(x)= e^{-\Phi(x)}T(x), \quad\quad \bar V(x) =T^{-1}(x)e^{-\Phi(x)} \label{V}
\b
This leads to the equivalent linear systems
\be
\left\{\brr{l}
(\partial_+ + A_+^{\mbox{\tiny V}})V\,=\,0
\\[10pt]
(\partial_- + A_-^{\mbox{\tiny V}})V\,=\,0
\err\right.\,,\qquad\qquad\mbox{\small with}\quad
\brr{l}
A_+^{\mbox{\tiny V}}\,=\,2\partial_+\Phi + \E_+ ~~~
\\[10pt]
A_-^{\mbox{\tiny V}}\,=\,e^{-2ad_\Phi}\E_-
\err
\label{Vconn}
\ee
and, similarly, for $\bar{V}$
\be
\left\{\brr{l}
\partial_+\bar{V} =\bar{V} A_+^{\bar{\mbox{\tiny V}}}
\\[10pt]
\partial_-\bar{V} =\bar{V} A_-^{\bar{\mbox{\tiny V}}}
\err\right.\,,\qquad\qquad\mbox{\small with}\quad
\brr{l}
A_+^{\bar{\mbox{\tiny V}}}\,=\,e^{2ad_\Phi}\E_+
\\[10pt]
A_-^{\bar{\mbox{\tiny V}}}\,=\,-2\partial_-\Phi + \E_-\,.
\err
\label{barVconn}
\ee

Notice that in terms of $V$ and $\bar V$, we have
\be
e^{-2\lambda^{(r)}(\Phi)}\,=\,<\Lambda^{(r)}|V \bar{V}
|\Lambda^{(r)}>\,=\,
\xi^{(r)}\cdot\,\bar{\xi}^{(r)}\,.
\0
\ee

We will see in the next sections that there are gauge transformations which map
$A_-^{\mbox{\tiny V}}$ to zero,
and other gauge transformations for the second system which
similarly maps $A_+^{\bar{\mbox{\tiny V}}}$ to zero. We obtain a system
which is similar in form to the DS system. However the two systems do not
yet coincide.  A further step is necessary: it
consists in diagonalizing the monodromy matrix $T$ -- actually this is
logically
the first operation one must carry out. After that we will be
able to reconstruct the true DS system. In the next section we will see
this method at work for the Liouville model.

\section{The Liouville model}

\subsection{Introduction}

The Liouville model is the $sl_2$ ToFT. In such a case we put
\a
\Phi(x) = \textstyle {1\over 2} \varphi(x) H, \quad\quad P(x)=p(x)H, \quad\quad
Q=qH,\quad\quad{\cal E}_\pm=E_\pm \0
\b
etc. In the defining representation, whose highest weight vector is $|\Lambda>$
=$\left( \matrix{1\cr 0}\right)$, we have
\a
H=\left(\matrix{1&0\cr 0&-1\cr}\right),
\quad\quad E_+=\left(\matrix{0&1\cr 0&0\cr}\right),
\quad\quad E_-=\left(\matrix{0&0\cr 1&0\cr}\right)\0
\b
The equation of motion is
\a
\partial_{+}\partial_{-}\varphi = e^{2\varphi}\label{liouville}
\b
We will also use the notation
\be
V\,=\,\left(\brr{cc} \xi_{11} & \xi_{12} \\
\xi_{21} & \xi_{22} \err\right),\quad\quad
\bar V\,=\,\left(\brr{cc} \bar \xi_{11} & \bar\xi_{12} \\
\bar \xi_{21} & \bar\xi_{22} \err\right),\quad\quad
T\,=\,\left(\brr{cc} \alpha& \beta\\
\gamma & \delta\err\right)\0
\ee

Our problem can be formulated as follows:

\noindent {\it For any solution $\varphi$ of the Liouville equation
construct a couple of real fields $p$ and $\bar p$ such that

1) p is chiral, $\bar p$ is antichiral;

2) $p$ and $\bar p$ are periodic;

3) they satisfy the Poisson brackets:
\a
\{p(x),p(y)\} &=& -(\partial_x -\partial_y) \delta (x-y) \0\\
\{p(x), \bar p(y)\} &=&0 \label{PB}\\
\{\bar p(x),\bar p(y)\} &=& (\partial_x -\partial_y) \delta (x-y) \0
\b
and such that the zero modes $ p_0$ and $\bar p_0$ coincide.
Moreover, construct $q$ such that
\a
\{q,p_0\}= \textstyle {1\over{\pi}}\label{PB0}
\b
and Poisson commute with the remaining degrees of freedom.
Verify that $p(x), \bar p(x)$ and $q$,
inserted into the DS construction give rise to exactly the solution $\varphi$
we started from.}

For pedagogical reasons in this section we will do everything in the
defining representation of $sl_2$. However the result extends to any
representation; we do not do it explicitly here since in the next section we
will provide a representation independent proof of the same results
for $sl_n$ Toda field theories.

\subsection{Reduction to the DS form}

Let us gauge away
$A_-^{\mbox{\tiny V}}$. We consider the transformation
\a
V\!\!\to\!\!h^{-1}V\,,
{}~~~~~~~~A_\pm^{\mbox{\tiny V}}\!\!\to\!\!^hA_\pm^{\mbox{\tiny V}}:=
h^{-1}A_\pm^{\mbox{\tiny V}}h+ h^{-1}\partial_\pm h\,,
\label{chir.gauge}
\b
where $h:=h(x,t)$ has the form
\[
h\,=\,\left(\brr{cc} 1 & 0 \\ a & 1 \err\right)\,.
\]
Imposing the condition we aim at, we find
\[
^hA_-^{\mbox{\tiny V}}\,=\,0\qquad\Rightarrow\qquad
\partial_-a + e^{2\varphi}\,=\,0
\]
In order to solve for $a$ we proceed as follows. We notice that
\be
a\,=\,-\partial_+ \varphi + (\mbox{\small\rm chiral terms})\,.\0
\ee
We can determine $a$ more precisely by imposing that $A_+^{\mbox{\tiny V}}$
remains upper triangular. Since
\a
h^{-1}V\,=\,\left(\brr{cc} \xi_{11} & \xi_{12} \\
-a\xi_{11} + \xi_{21} & -a\xi_{12} + \xi_{22} \err\right)\0
\b
we must have
\a
-a\xi_{11} + \xi_{21}=0\0
\b
In this way we find
\be
A_+^{({\sss +})}\,:=\,^gA_+^{\mbox{\tiny V}}=
\left(\brr{cc} -p & 1 \\
0& ~~p \err\right)\,,
\label{riccati}
\ee
where
\be
p\,=\,-\partial_+\varphi-\xi_{21}\xi_{11}^{-1}\,.
\label{chir.p}
\ee
It is easy to verify by a direct computation that $\partial_-p=0$.

We can do the same for the antichiral sector using a strictly upper triangular
matrix $\bar h$, and find
\be
^{\bar h}A_+^{\bar {\mbox{\tiny V}}}=0,\quad\quad
A_-^{({\sss -})}\,:=\,^{\bar h}A_-^{\bar {\mbox{\tiny V}}}=
\left(\brr{cc} -\bar p & 0 \\ 1 & \bar p \err\right)\,,
\label{achir.conn}
\ee
where
\be
\bar p\,=\,\partial_-\varphi-\bar\xi_{12}\bar\xi_{11}^{-1}\,.
\label{achir.p}
\ee
is antichiral.

It looks like we have reached our goal, that is we have reconstructed the
DS system, but it is not so. For, although
the Poisson bracket of $p$ with $p$ and $\bar p$ with $\bar p$ are the
expected ones (\ref{PB}), $p$ and $\bar p$ do not Poisson commute.
Moreover, by considering the effect of the monodromy matrix, we can quickly
verify that $p$ and $\bar p$ are not periodic. To obtain the correct answer
we have first to diagonalize the monodromy matrix (the further step
announced above).

\subsection{Diagonalizing the monodromy}.

In section 2 we introduced the initial condition $T(0)=1$.
However there is no a priori reason why we should use precisely this condition.
We are free to change it at will. We take advantage of this freedom to
diagonalize the monodromy matrix. More precisely,
we change the initial data of the linear systems by making the transformation
\[
V\longrightarrow V\, g\;\; ,\;\;\;\;
{\bar V}\longrightarrow g^{-1}{\bar V}\,
\]
with $g$ a unimodular $2\times 2$ constant matrix. Accordingly, the monodromy
will change as
\[
T\longrightarrow T^{g}=g^{-1}Tg
\]
We would like $T^{g}$ to be diagonal
(or, at least, triangular). This is always possible over the complex numbers,
i.e. with $g\in \Slc$.
However, given a real solution $\varphi$ of the Liouville equation,
we are interested in doing this over the
reals, i.e. with $g \in \Slr$, since only in this case will the $p$
and $\bar p$ fields be real
(see below). Although the answer to this question is known to be positive
\cite{FT},
we have not been able to find an explicit proof of it in the literature.
Therefore we think it useful, also in view of the $sl_n$ case,
to exhibit an explicit proof that for regular real solutions $\varphi$ the
monodromy is hyperbolic, i.e. $tr~T>2$.
The proof is given in Appendix A; here we assume the result, which allows us
to conclude that $T$ can be diagonalized by means of a real $g$ matrix.

For the diagonalizing matrix $g$ let us make the position
\[
g=R\, S
\]
with
\[
R=\left(\brr{cc}1&0\\ r&1\err\right)\, ,\;\;\;\;
S=\left(\brr{cc}1&s\\ 0&1\err\right)
\]
In this way $r$ satisfies a second
order equation and $s$ is determined after $r.$ We could have made
other choices for $R$ and $S$, the results would be completely equivalent.
What is
important is that in the hyperbolic case we have two real solutions for
$r,$ namely
\[
r_\pm = \frac{-(\alpha - \delta )\pm \sqrt{\Delta}}{2\beta}
\]
where
\[
\Delta =(\alpha +\delta )^2-4\equiv (\mbox{tr}\,T)^2-4
\]
is the positive discriminant. The corresponding elements $s_\pm$ are given by
\[
s_\pm =\mp \frac{\beta}{\sqrt{\Delta}}
\]

Thus we distinguish between the two solutions by appending a plus or
minus sign: $g_\pm .~$ Their explicit forms are
\[
g_\pm=
\left( \brr{cc}1&\mp \frac{\beta}{\sqrt{\Delta}}\\
r_\pm &\mp \frac{\beta}{\sqrt{\Delta}}r_\mp\err\right)\,\0
\]
The difference between the two is that ${g_+}^{-1}Tg_+=D$,
$~D=\mbox{diag}(\lambda_+,\lambda_-)$, while
${g_-}^{-1}Tg_-={D^{-1}}$,
where $\lambda_\pm$ are
the eigenvalues of $T.$ We will clarify later on the significance of these two
possibilities. Henceforth, when the labels $\pm$ are omitted we mean that
the equations or statements are true for both choices, i.e. both for +
and for --.

After this lengthy discussion of the diagonalizing matrix $g$, let us return
to the main problem. Let us apply the same procedure as in the last subsection,
starting from $Vg$ and $g^{-1}\bar V $ instead of $V$ and $\bar V$, i.e.
we determine transformations $h_g$ and $\bar h_g$ which play the role of
$g$ and $\bar g$, respectively. We end up with
\a
^{h_g}A_-^{\mbox{\tiny Vg}}\,=\,0,\quad\quad\quad
\hat A_+\,:=\,^{h_g}A_+^{\mbox{\tiny Vg}}\,=\,
\left(\brr{cc} -p & 1 \\ 0 & p \err\right)\label{^A+}
\b
and
\a
^{\bar h_g}A_+^{\bar {\mbox{\tiny V}}g}\,=\,0,\quad\quad\quad
\hat A_-\,:=\,^{\bar h_g}A_+^{\bar{\mbox{\tiny V}}g}\,=\,
\left(\brr{cc} -\bar p & 0 \\ 1 & \bar p \err\right)\label{^A-}
\b
where $p$ and $\bar p$
are represented by the bosonization formulas
\ba
p\!\!&=&\!\! \partial_x\log \xi_{11}^{g}\,=
\!\partial_x \log \left(\xi_{11}+ r \xi_{12}\right)\,,
\label{chir.bos.2}\\
\bar p\!\!&=&\!\! -\partial_x\log \bar \xi_{11}^{g}\,=
\!-\partial_x \log\left( (1+rs)\,\bar\xi_{11} - s \bar\xi_{21}\right)\,.
\label{achir.bos.2}
\ea
Here $\xi^g_{ij}$ ($\bar\xi^g_{ij}$) represent the elements of the matrix
$Vg$ ($\bar Vg$). Remember that each of the above equations is
a shorthand for a pair of equations corresponding to the two solutions
$g_\pm$, so that we have in fact the pair $p_+, \bar p_+$ and
the pair $p_-, \bar p_-$. For both choices,
i.e. either + and --, $p$ and $\bar p$ are now periodic fields which satisfy
the Poisson brackets (\ref{PB})\footnote{The Poisson brackets are calculated
starting from the brackets
(\ref{Txi})and those given in Appendix C}.

We notice that the zero modes of the fields $p$ and $\bar p$ are well defined.
Indeed
\bean
p_{\pm ,0}&=&\log\lambda_\pm\\
\bar p_{\pm ,0}&=&-\log\lambda_\mp
\eean
So, as expected, they are equal, $p_0=\bar p_0$ (indeed
$\lambda_-=1/\lambda_+$).

Summarizing, we have (almost) achieved to reconstruct the DS linear system
in terms of the interacting field $\varphi$, since
\a
&&(\partial_+ +\hat A_+){\cal V}^g_+=0, \quad\quad \partial_- {\cal
V}^g_+=0\0\\
&&\partial_+ \bar{\cal V}^g_-=0,\quad\quad
\partial_-{\cal V}^g_- -{\cal V}^g_-\hat A_-=0\0
\b
where $\hat A_\pm$ are given above and
\a
{\cal V}^g_+=h_g^{-1}Vg,\quad\quad\quad {\cal V}^g_-=g^{-1}\bar V\bar h_g
\label{calV}
\b
Moreover
\a
e^{-\varphi}\,=\,<\Lambda|{\cal V}^g_+\,{{\cal V}}^g_-|\Lambda >\label{recon1}
\b
However there are still two undefined points. The first is that
${\cal V}^g_+$ and
${\cal V}^g_-$ cannot yet be identified with ${\cal Q}_+$ and
${\cal Q}_-$ as the initial
conditions are different. The second point is that we have not yet
retrieved the conjugate variable $q$ to the zero modes (see Remark 1 at the end
of the previous section). The two points are related.

Let us recall eqs.(\ref{recon1}) and the reconstruction formula for the
solutions of the previous section, and notice that
if we want the DS systems to coincide the
relation among them must be
\[
{\cal V}^g_+\,{\cal V}^g_-={\cal Q}_+g_S\,{\cal D}\,\bar g_S^{
-1}\, {\cal Q}_-
\]
The simplest thing to do is to examine this relation at the origin,
where ${\cal Q}_+(0)={\cal Q}_-(0)=1.$ So we must have
\a
{\cal V}^g_+(0){\cal V}^g_-(0)= g_S {\cal D}\bar g_S^{-1}\label{0}
\b
On the other hand, due to the normalization
condition $T (0)=1$ on the transport matrix, at the origin we have
$V(0)=e^{-\Phi (0)}$ and $\bar V(0)=e^{-\Phi (0)}$.
After some algebra one finds
\[
{\cal V}^g_+(0)=\left(\brr{cc}e^{-\varphi_0/2}&s\, e^{-\varphi_0/2}\\
0&e^{\varphi_0/2}\err\right)\,\;\;\;\;
\bar {\cal V}^g_-(0)=\left(\brr{cc}(1+sr)e^{-\varphi_0/2}&0\\
-re^{-\varphi_0/2}&(1+sr)^{-1}e^{\varphi_0/2}\err\right)
\]
where $\varphi_0=\varphi(0)$.
Now the matrices $g_S$ and $\bar g_S$
are unipotent, hence they can be uniquely determined on the basis of
eq.(\ref{0}). Simultaneously the matrix ${\cal D}$ is uniquely identified
\[
{\cal D} =\left(\brr{cc}(1+sr)e^{-\varphi_0}&0\\
0&(1+sr)^{-1}e^{\varphi_0}\err\right)
\]
One can verify that
\be
\{p_0,{\cal D}\}=\frac{1}{\pi}H{\cal D}\label{rho}
\ee
which is what we wanted to prove.

In conclusion, we have reconstructed all the elements of the DS linear system
in terms of the periodic solution $\varphi$ of the Liouville equation.

\subsection{Action of the Weyl group}.

Let us return now to the meaning of the existence
of the two pairs $p_+, \bar p_+$ and $p_-, \bar p_-$ of free fields
satisfying all our
requirements. First we notice that we can pass from the diagonal
matrix $D$ to $D^{-1}$ via a Weyl transformation
\[
D^{-1}=w D w^{-1}\, ,\;\;\;\;
w=\left(\brr{cr}0&-1\\1&0\err\right)
\]
(we identify the group element $w$ with the element of the Weyl group
it represents). It follows that $g_-w g_+^{-1}$ commutes with $T,$ so
that $z=g_+^{-1}g_-w$ commutes with $D,$ hence is diagonal.

By construction, under a monodromy transformation we have
\bean
&&{\cal V}^{g_+}_+\longrightarrow {\cal V}^{g_+}_+ D,\quad\quad
{\cal V}^{g_+}_-\longrightarrow D^{-1}{\cal V}^{g_+}_- \\
&&{\cal V}^{g_-}_+\longrightarrow {\cal V}^{g_-}_+ D^{-1},\quad\quad
{\cal V}^{g_-}_-\longrightarrow D{\cal V}^{g_-}_- \\
\eean
but we also have
\[
{\cal V}^{g_-}_+w\,\longrightarrow {\cal V}_+^{g_-}w D\, ,\;\;\;\;
w^{-1}{\cal V}^{g_-}_-\longrightarrow D^{-1}w^{-1}{\cal V}^{g_-}_-\,
\]
so that the quantities ${\cal V}_+^{g_-}w$ and
$w^{-1}{\cal V}_-^{g_-}$ have the same monodromy as
${\cal V}_+^{g_+}$\,  and \,${\cal V}_-^{g_+}$, respectively. Actually we have
${\cal V}_+^{g_-}w={\cal V}_+^{g_-} z$ and
$w^{-1}{\cal V}_-^{g_-}= z^{-1}{\cal V}_-^{g_+} $ and since $z$ is diagonal,
the free bosonic fields constructed from ${\cal V}_+^{g_-}w$
and $w^{-1}{\cal V}_-^{g_-}$
are in fact equal to $p_+$ and $\bar p_+.$  A similar conclusion holds for
$p_-$ and $\bar p_-$.

The outcome of this discussion is that we can pass from one pair of
solutions $p,\bar p$ to the other pair through the action of the Weyl group.

\section{The $sl_n$ Toda field theories}

In this section we generalize to $sl_n$ Toda field theories what we
have done in the previous section for the Liouville theory.
In more detail, we do the following:

\noindent {\it For any hyperbolic solution $\Phi$ of the Toda field equations
we construct a couple of real fields $P$ and $\bar P$ valued in the Cartan
subalgebra such that

1) $P$ is chiral, $\bar P$ is antichiral;

2) $P$ and $\bar P$ are periodic;

3) they satisfy the Poisson brackets (\ref{PP}).

Moreover the zero modes $ P_0$ and $\bar P_0$ coincide and we construct
the conjugate of them. $P$ and $\bar P$, inserted in the DS
construction, give back the solution we started from. We also show that there
are many possible choices of the pair $P$ and $\bar P$ for any given solution,
such choices being mapped into one another by the action of the Weyl group.}

\noindent Two remarks are in order:

-- Contrary to the previous section, what we do in this section is
explicitly representation independent.

-- A large part of this section applies to Toda field theories based on any
simple Lie algebra; however the proof is tailored to $sl_n$ and further work
is needed to extend the results to any simple Lie algebra.

Let us pass now to the proofs. Instructed by the $sl_2$ case, we proceed
as follows. First we prove that the monodromy matrix is hyperbolic (this is
actually done in Appendix B). Then we apply the Cartan decomposition to the
transport matrix and extract the $P$ and $\bar P$ fields. Next we prove that
they have the correct Poisson brackets (\ref{PP}). Finally we discuss the
zero modes and their conjugate momenta. As an aside we obtain the result
concerning the action of the Weyl group.

\subsection{Diagonalizing the Monodromy Matrix}

In Appendix B we prove that the monodromy matrix $T$ for hyperbolic
solutions of any $\G=sl_n$
Toda field theory is hyperbolic, i.e. has real positive eigenvalues and
can be diagonalized within $\Slnr$. I.e. we have
\a
T^g&=&g^{-1}Tg\label{diag'}
\b
where $T^g$ is diagonal matrix with positive diagonal entries and $g$ is
a suitable element of $\Slnr$. Here we assume this result and show
some simple consequences.

Let us consider an arbitrary finite dimensional representation of the Lie
algebra $\G$, which can be lifted to a representation of $\Slnr$.
The representation space is spanned by
vectors $|\mu , m>$ where $\mu $ is a weight and $m$ counts its
degeneracy. We have
\a
T^g |\mu , m>&=& \mu (T^g) |\mu , m>\label{diag''}
\b

The first consequence we extract from eqs.(\ref{diag'},\ref{diag''})
is the Poisson brackets of an arbitrary dynamical variable $\A$, $\{ \A , g\}$,
in terms of simpler and known Poisson brackets.
We easily find
\a
<\nu ,n|g^{-1} \{ \A , g\} |\mu , m>&=&
\frac{1}{ \mu (T^g)-\nu (T^g)}
<\nu ,n|g^{-1} \{ \A, T \} g |\mu , m>\0\\
\mu &\neq &\nu
\label{basic}
\b
This identity fixes $g^{-1} \{ \A , g\}$ up to
terms belonging to the Cartan subalgebra. In fact, in order to find
$\{ P\x P\}, \, \, \, \{\bar{P}\x \bar{P}\}$ and
$\{P \x \bar{P}\}$ we do not need to know explicitly these
terms; however they are important in showing that the fields
$P$ and $\bar{P}$ have the correct Poisson brackets with
the conjugate variable $\D$. In order to find a more complete result
let us recall the Cartan decomposition of $\G$
\a
\G=\N_+ \oplus \H \oplus \N_-
\b
and its Borel subalgebras
\a
\B_{\pm}=\N_{\pm} \oplus \H
\b
By $\P_\H, \, \P_{\N_{\pm}}$ we will denote the
projection operators on the corresponding subalgebras.

Next, our first observation is that if
$g=g_- g_+ k$ ( $g_{\pm}\in e^{\N_{\pm}}, \, \, \,
k\in e^{\H }$ ) diagonalizes the monodromy matrix, the
element $g k^{-1}$ will also diagonalize $T$. Therefore,
we can set $g=g_+g_-$. For this special choice of the diagonalizing element
the Poisson bracket
\a
g^{-1}_-\cdot \{ \A , g \}\cdot g^{-1}_+ \0
\b
is orthogonal to the Cartan subalgebra $\H$. From this observation
and from (\ref{basic}) one obtains
\a
 g^{-1} \{ \A , g\}&=&
- \sum_{ \stackrel{\stackrel{\mu \neq \nu}{m,n}}{\sss\al>0}}\sum_{p=1}^{n-1}
\frac{1}{ \mu (T^g)-\nu (T^g)}\cdot    \label{Ag}\\
&&\hskip -2.5 cm
\cdot\frac{ <\nu ,n|g^{-1} \{ \A, T \} g |\mu , m>
<\mu , m| E_{\al}|\nu ,n>(g_+ E_{-\al} g_+^{-1}\, , \, H_p)}
{(E_{\al}\, , \, E_{-\al})} \cdot H_p+\0\\
&&\hskip -2.5 cm + \sum_{ \stackrel{\stackrel{\mu \neq \nu}{m,n}}{\sss\al}}
\frac{1}{ \mu (T^g)-\nu (T^g)}\cdot
\frac{ <\nu ,n|g^{-1} \{ \A, T \} g |\mu , m>
<\mu , m| E_{-\al}|\nu ,n>}{(E_{\al}\, , \, E_{-\al})}
\cdot E_{\al} \0
\b
where $\al$ in the first term runs over the set of all positive
roots while the summation in $\al$ in the second term is
over all (positive and negative) roots.

We shall finish this subsection with the following remark.
The matrix $g$ which diagonalizes the monodromy matrix $T$
is not fixed uniquely. This comes from the fact that the
Cartan subgroup is invariant under the elements
$kw$ where $k$ is in the Cartan subgroup and $w$ belongs
to the Weyl group $\W$ of the Lie group $G$. The later
is defined as the  quotient $N(H)/H$, $N(H)$ being the normalizer
of the Cartan subgroup $H$. For $SL(n)$ the Weyl group is isomorphic to
the permutation group $\S_n$ of $n$ elements.

\subsection{Gauss decomposition and $P$, $\bar P$ fields.}

Let $g$ be one of the diagonalizing elements introduced in the previous
subsection. The  transport matrices
\a
V^g&=&Vg \0\\
\bar{V}^g&=&g^{-1} \bar{V}\0
\b
satisfy the same linear systems as $V$ and $\bar{V}$
and have diagonal monodromy
\a
V^g(x+2\pi)&=&V(x) T^g  \0\\
\bar{V}^g(x+2\pi )&=& (T^g)^{-1} \bar{V}^g(x)\0\\
T^g&=&g^{-1} T g\0
\b
{}From the Gauss decomposition
\a
V^g(x)&=&N^g_-(x)e^{K^g(x)}M_+^g(x)\0\\
\bar{V}^g(x)&=&\bar{N}_-^g(x) e^{\bar{K}^g(x)} \bar{M}_+^g(x)
\label{Gauss}
\b
where $ N_-^g  , \bar{N}_-^g \in e^{ \N_-};
\, \, \, K^g  ,  \bar{K}^g \in \H; \, \, \,
M^g_+ , \bar{M}^g_+ \in   e^{ \N_-} $, and from the
linear systems (\ref{LS}) one obtains that the matrices
\a
\V_+^g&=&e^{K^g(x)}M_+^g(x)\0\\
\V_-^g&=&\bar{N}_-^g(x) e^{\bar{K}^g(x)}
\label{vu}
\b
satisfy the DS systems (\ref{DS+}) and (\ref{DS-}) with
\a
P^g(x)&=& \partial_x K^g(x)\0\\
\bar{P}^g(x)&=& \partial_x \bar{K^g}(x)
\label{campi}
\b
where, out of clarity, we have explicitly denoted the dependence on the
diagonalizing element $g$. In the following we will at times understand
the label $g$ in $P$ and $\bar P$.

\subsection{Poisson brackets of $P$ and $\bar P$}.

In the previous subsection we have defined the fields $P$ and $\bar P$.
Now we prove that their Poisson on brackets are exactly (\ref{PP}).
This subsection is rather technical, but we think it instructive
to show explicitly at least in this case what kind of computations
are involved.

{}From (\ref{Gauss}) it follows that $\{ P^g \x P^g \}$
coincides with the projection on $\H\o \H$ of
\a
\partial_x \partial_y \left(N_-^g(x)^{-1}\o N_-^g(y)^{-1} \cdot
\{ V^g(x) \x V^g(y) \} \cdot \V_+^g(x)^{-1}
\o  \V_+^g(y)^{-1}\right)
\label{pp1}
\b

We will consider separately the various contributions
to $\{ V^g \x V^g \}$. One such contribution is
given by $\{ V \x V \} \cdot g\o g$ and
produces the terms
\a
\P_\H \o \P_\H \cdot
\left( Ad\left(\V_+^g(x)\right)\o Ad\left(\V_+^g(y)\right)
\left( \theta (x-y) (r^+)^g +\theta (y-x) (r^-)^g \right) \right)&=& \0\\
\hskip -7 cm \P_\H \o \P_\H
\left( \theta (x-y) (r^+)^g +\theta (y-x) (r^-)^g \right) +\0\\
\hskip -7 cm+\P_\H Ad\left(\V_+^g(x)\right)\P_{\N_-}
\o \P_\H \left( r^g \right)
 +\P_\H\o \P_\H
Ad\left(\V_+^g(y)\right)\P_{\N_-}
\left( r^g \right) +\0\\
\hskip -7 cm +\P_\H Ad\left(\V_+^g(x)\right)\P_{\N_-}
\o \P_\H Ad\left(\V_+^g(x)\right)\P_{\N_-}
 \left(r^g \right)\0
\b
where
\a
Ad\left(\V_+^g(x)\right) (a)&=&\V_+^g(x) a \V^g_+(x)^{-1}, \quad\quad\quad
a  \in  {\cal G}\0\\
(r^{\pm })^g&=& Ad (g^{-1})\o  Ad (g^{-1}) (r^{\pm })\0
\b

Taking into account that
\a
(r^+ )^g-(r^-)^g=r^+-r^-\0
\b
we obtain the following identities
\a
\P_\H \o \P_\H (r^+)^g&=&
\P_\H \o \P_\H (r^-)^g+
2\sum_i H_i \o H_i \0\\
\P_\H \o \P_{\N_{\pm}} (r^+)^g&=&
\P_\H \o \P_{\N_{\pm}} (r^-)^g\0\\
\P_{\N_{\pm}} \o \P_\H (r^+)^g&=&
\P_{\N_{\pm}} \o \P_\H (r^-)^g\0
\b
from which it then follows that the corresponding contribution
to $\{ P^g \x P^g \} $ is reduced to the expression
\a
\partial_x \partial_y \epsilon (x-y) \cdot t_0+
\partial_x \partial_y \P_\H Ad\left(\V_+^g(x)\right)\P_{\N_-}
\o \P_\H Ad \left(\V_+^g(y)\right)\P_{\N_-}
 \left(r^g \right)
\label{primo}
\b
while the other two terms give vanishing contributions to this Poisson
bracket since they depend only either on $x$ or on $y$.

Next we consider the other two contributions
\a
1\o V(y) \cdot \{ V(x) \x g \} \cdot g \o 1+
V(x) \o 1 \cdot \{ g \x V(y) \} \cdot 1 \o g\label{Vg+gV}
\b
to $\{ V^g \x V^g \}$. We remark that
\a
Ad \left(\V^g (x)\right)(h)=h, \quad\quad\quad {\rm if}\quad h\in  \H\0
\b
Therefore the terms belonging to $\H$ of the
Poisson brackets (\ref{Vg+gV}) do not produce contributions to
$\{ P^g \x P^g\}$. From this and from the main result (\ref{Ag})
 of the subsection 4.1, we obtain that (\ref{Vg+gV})
gives the contribution
\a
\hskip -2 cm \sum_{ \stackrel{\stackrel{\l \neq \s}{l,s}}{\sss\alpha>0}}
\sum_{ \stackrel{\stackrel{\mu \neq \nu}{m,n}}{\sss\beta>0}}
&&\left(\frac{\nu(T^g)}{ \mu (T^g)-\nu (T^g)}+
\frac{\s (T^g)}{\l (T^g)-\s (T^g)} \right) \cdot \0\\
&&\hskip - 2.5 cm \cdot
\frac{<\l , l| E_{\al}|\s ,s><\mu , m| E_{\beta}|\nu ,n>}
{(E_{\al}, E_{-\al})(E_{\beta}, E_{-\beta})}
<\s , s|\o\!\! < \nu , n | (r)^g |\l ,l>\!\! \o |\mu ,m>\cdot \0
\b
\a
&&\hskip -2 cm \cdot
\partial_x \partial_y {\P}_{\H} Ad \left(\V^g_+(x)\right)
\o {\P}_{\H} Ad \left(\V^g_+(y)\right)
\left( E_{-\al} \o E_{-\beta} \right)
\label{secondo}
\b
Finally the term
\a
V(x) \o V(y) \cdot \{ g \x g \} \0
\b
gives the contribution
\a
\hskip -2 cm \sum_{ \stackrel{\stackrel{\l \neq \s}{l,s}}{\sss\al>0}}
\sum_{ \stackrel{\stackrel{\mu \neq \nu}{m,n}}{\sss\beta>0}}
&&\frac{\nu(T^g)\s (T^g)-\l (T^g)\mu (T^g)}
{(\mu(T^g)-\nu(T^g))(\l (T^g)-\s (T^g))} \cdot\0\\
&&\hskip - 2.5 cm \cdot
\frac{<\l , l| E_{\al}|\s ,s><\mu , m| E_{\beta}|\nu ,n>}
{(E_{\al}, E_{-\al}) (E_{\beta}, E_{-\beta})}
<\s , s|\o \!\! < \nu , n | (r)^g |\l ,l>\!\! \o |\mu ,m>\cdot\0
\b
\a
&&\hskip -2 cm \cdot
\partial_x \partial_y {\P}_{\H} Ad\left( \V^g_+(x)\right)
\o {\P}_{\H} Ad \left(\V^g_+(y)\right)
\left( E_{-\al} \o E_{-\beta} \right)
\label{terzo}
\b
{}From the identity
\a
\hskip -1 cm \frac{\nu(T^g)}{ \mu (T^g)-\nu (T^g)}+
\frac{\s (T^g)}{\l (T^g)-\s (T^g)}+
\frac{\nu(T^g)\s (T^g)-\l (T^g)\mu (T^g)}
{(\mu(T^g)-\nu(T^g))(\l (T^g)-\s (T^g))}=-1
\label{trucco}
\b
it follows that (\ref{secondo}) and (\ref{terzo}) cancel
the second term of (\ref{primo}) and therefore
\a
\{ P^g(x) \x P^g(y) \}&=&
\partial_x \partial_y \epsilon (x-y) \cdot t_0 \0
\b
Analogously in the antichiral
sector we obtain
\a
\{ \bar{P}^{\bar g}(x) \x \bar{P}^{\bar g}(y) \}&=&
-\partial_x \partial_y \epsilon (x-y) \cdot t_0 \0
\b
where $\bar g$ is another diagonalizing element, in general different from
$g$.

Next we show that the Poisson bracket
$\{ P^g \x \bar{P}^{\bar{g}}\} $ vanishes. Proceeding
as in the previous subsection we first obtain that the contribution of the term
\a
1\o \bar{g}^{-1} \cdot \{ V(x) \x \bar{V}(y) \}
g\o 1\0
\b
is given by
\a
-\partial_x \partial_y
\P_{\H} Ad\left(\V^g_+ (x)\right) \P_{\N_-} \o
\P_{\H} Ad \left( \V^{\bar{g}}_-(y)^{-1}\right) \P_{\N_+}
\left( (r^-)^{g \bar{g}}\right)
\label{uno}
\b
where
\a
(r^-)^{g \bar{g}}=g^{-1}\o\bar{g}^{-1} \cdot  r^-
g\o \bar{g}\0
\b
The contribution of
\a
V(x)\o \bar{g}^{-1} \cdot \{ g \x \bar{V}(y) \}+
\{ V(x) \x \bar{g}^{-1} \} \cdot g \o \bar{V}(y)\0
\b
reads
\a
\hskip -2 cm - \sum_{ \stackrel{\stackrel{\l \neq \s}{l,s}}{\sss\al>0}}
\sum_{ \stackrel{\stackrel{\mu \neq \nu}{m,n}}{\sss\beta>0}}
&&\left(\frac{\nu(T^{\bar{g}})}{ \mu (T^{\bar{g}})-\nu (T^{\bar{g}})}+
\frac{\s (T^g)}{\l (T^g)-\s (T^g)} \right) \cdot \0\\
&&\hskip - 2.5 cm \cdot
\frac{<\l , l| E_{\al}|\s ,s><\mu , m| E_{-\beta}|\nu ,n>}
{(E_{\al}, E_{-\al}) (E_{\beta}, E_{-\beta})}
<\s , s|\o \!\! < \nu , n |
(r^-)^{g \bar{g}} |\l ,l>\!\! \o |\mu ,m>\cdot\0
\b
\a
&&\hskip -2 cm \cdot
\partial_x \partial_y {\P}_{\H} Ad \left(\V^g_+(x)\right) \o
{\P}_{\H} Ad \left(\V^{\bar{g}}_-(y)^{-1}\right)
\left( E_{-\al} \o E_{\beta} \right)
\label{due}
\b
and the term
\a
V(x)\o 1 \cdot \{ g\x \bar{g}^{-1} \} \cdot 1\o \bar{V}(y)\0
\b
produces
\a
\hskip -2 cm -\sum_{ \stackrel{\stackrel{\l \neq \s}{l,s}}{\al>0}}
\sum_{ \stackrel{\stackrel{\mu \neq \nu}{m,n}}{\beta>0}}
&&\frac{\nu(T^{\bar{g}})\s (T^g)-\l (T^g)\mu (T^{\bar{g}})}
{(\mu(T^{\bar{g}})-\nu(T^{\bar{g}}))(\l (T^g)-\s (T^g))} \cdot\0\\
&&\hskip - 2.5 cm \cdot
\frac{<\l , l| E_{\al}|\s ,s><\mu , m| E_{-\beta}|\nu ,n>}
{(E_{\al}, E_{-\al}) (E_{\beta}, E_{-\beta})}
<\s , s|\o \!\! < \nu , n |
(r^-)^{g \bar{g}} |\l ,l>\!\! \o |\mu ,m>\cdot\0
\b
\a
&&\hskip -2 cm \cdot
\partial_x \partial_y {\P}_{\H} Ad \left(\V^g_+(x)\right) \o
{\P}_{\H} Ad \left(\V^{\bar{g}}_+(y)^{-1} \right)
\left( E_{-\al} \o E_{\beta} \right)
\label{tre}
\b
Using again (\ref{trucco}) we conclude that $P^g$
and $\bar{P}^{\bar{g}}$ Poisson commute.

\subsection{The variable conjugate to the zero modes.}

Let us choose now a fixed diagonalizing element $g$ for both the chiral and
anti--chiral part. Then, by construction, the zero modes $P^g_0$ and
$\bar P^g_0$ coincide. It remains for us to reconstruct the conjugate variable
to these zero modes.

Let $|\Lambda>$ be the highest weight vector of the representation we choose to
work on. The Gauss decomposition (\ref{Gauss}) allow us to express
the Toda fields as
\a
e^{-2\lambda (\Phi)}&=&<\Lambda|V(x)\bar{V}(x)|\Lambda>=
<\Lambda|V^g(x)\bar{V}^g(x)|\Lambda>=\0\\
&=& <\Lambda|\V_+^g(x)\V_-^g(x)|\Lambda>=<\Lambda|\Q_+^g(x)
\V_+^g(0)\V_-^g(0)\Q_-^g(x)|\Lambda>\label{reconstruction}
\b
where the chiral objects
\a
\Q_+^g(x)&=&\V_+^g(x)\V_+^g(0)^{-1}\0\\
\Q_-^g(x)&=&\V_-^g(0)^{-1} \V_-^g(x)
\b
satisfy the DS systems (\ref{DS+}), (\ref{DS-}).
The Gauss decomposition of the constant matrix
\a
\V_+^g(0)\V_-^g(0)=g_S D \bar{g}_S^{-1}, ~~~~~
g_S\in e^{\N_{+}}, ~~~~~  \bar{g}_S\in e^{\N_-}, ~~~~~ \D\in e^{\H}
\b
is obtained from (\ref{Gauss})
\a
g_S=e^{K^g(0)} M_+^g(0) e^{-K^g(0)}, ~~~~~~
\bar{g}_S=e^{-\bar{K}^g(0)} \bar{N}_-^g(0)^{-1} e^{\bar{K}^g(0)},~~~~~~
D=e^{K^g(0)+ \bar{K}^g(0)}\0
\b
A calculation a bit involved but straightforward shows
that
\a
\{ P(x)\x D\}&=& 2\delta(x) \sum_i H_i\o H_i D \0\\
\{ \bar{P}(x) \x D\}&=& 2\delta(x) \sum_i H_i\o H_I D
\b

\subsection{The action of the Weyl group}.

We have already noted that the element which diagonalizes
the monodromy matrix is not uniquely fixed. Therefore, as we have done in
subsection 4.3,
one can choose two different diagonalizing elements
$g, \, \, \,\bar{g}\in \Slnr$  and set
\a
V^g&=&V g,\quad\quad\quad
\bar {V}^{\bar{g}}=\bar{g}^{-1} \bar {V},\quad\quad\quad
g\cdot \bar{g}^{-1} \in e^{\H}\cdot \W\0
\b
The corresponding free chiral fields (\ref{campi})
$P^g$ and $\bar{P}^{\bar{g}}$ are obviously periodic.
On the other hand the right (left)
multiplication of $V$ ($\bar{V}$) by elements of the Cartan
subgroup shifts the fields $K^g$ ($\bar{K^g}$) (see (\ref{Gauss}),
(\ref{vu}) )
by a constant and
thus does not change the momenta $P$ ($\bar{P}$). This observation
allows us to state that only the Weyl group acts nontrivially on
these fields.

Therefore there are at least $|\W|^2$ possibilities
to construct a pair of free chiral fields $(P,\bar{P})$
with the right Poisson brackets. However the number of possible
choices reduces to $|\W|$ if one
wants, as we do, the Toda field to be reconstructed according to the formula
(\ref{reconstruction}).
In fact this formula implies that $g\cdot \bar{g}^{-1}
\in e^{\H}$ and therefore, without changing $P$ and $\bar{P}$
one can set $g\cdot \bar{g}^{-1}=1$.

\section{Parabolic monodromies.}

In this section we consider other types of non--regular solutions of the
Liouville equation. They correspond to parabolic
monodromies. The following discussion does not pretend to cope with the problem
of classifying all singular solutions. But it is nevertheless interesting
to see how a large class of them can be represented by means of free fields.

We come across these singular solutions in the following way. Let us return to
subsection 3.2. There we found two fields, a chiral field $p$, (\ref{chir.p}),
and an antichiral one $\bar p$, (\ref{achir.p}), which were almost the correct
solution to our problem but not quite, in that they do not Poisson commute.
For convenience we relabel them $p_1$ and $\bar p_1$. Now instead of
proceeding,
as we did in section 3, with the diagonalization of the monodromy, we can
remark
that there is a gauge freedom left after gauging away $A_-^{\mbox{\tiny V}}$
and $A_+^{\bar {\mbox{\tiny V}}}$. We can take advantage of this in order to
modify $p_1$ and $\bar p_1$. It turns out, for example, that we can
add to $p_1$ an object of weight 1 which is a total derivative of $x_+$.
After some calculations one realizes that there are two other interesting
chiral and antichiral fields, precisely
\bean
p_2=-(\dpiu\varphi +\xi_{22}/\xi_{12}), \quad\quad\quad
\bar p_2=-\dmeno\varphi +\bar\xi_{22}/\bar\xi_{21})
\eean
We notice that these can also be obtained from the previous ones
via the action of the Weyl group.

One then computes all the possible Poisson brackets among the
elements of these two pairs and we finds the following
result
{\em The pairs $(p_1\, ,\,\bar p_2)\, ,$ $(p_2\, ,\,\bar p_1)\, ,$ and
$(p_2\, ,\,\bar p_2)\, ,$ all satisfy the Poisson brackets (\ref{PB}
\/}).

This is of course only an intermediate result, since we must address
the question of periodicity with the additional problem of
deciding which choice is the most suitable one for our purposes. In
order to discuss periodicity, we consider the effect of a monodromy operation
\bean
\xi_{11}\rightarrow\alpha\xi_{11} +\gamma \xi_{12}\;\;\;\;& &\;\;\;\;
\bar\xi_{11} \rightarrow\delta\bar\xi_{11} -\beta \bar\xi_{21}\\
\xi_{12} \rightarrow \beta\xi_{11} + \delta \xi_{12}\;\;\;\;& &\;\;\;\;
\bar\xi_{12} \rightarrow \delta\bar\xi_{12} - \beta \bar\xi_{22}\\
\xi_{21} \rightarrow \alpha\xi_{21} +\gamma \xi_{22}\;\;\;\;& &\;\;\;\;
\bar\xi_{21} \rightarrow -\gamma\bar\xi_{11} +\alpha\bar\xi_{21}\\
\xi_{22} \rightarrow \beta \xi_{21} + \delta \xi_{22}\;\;\;\;& &\;\;\;\;
\bar\xi_{22} \rightarrow -\gamma \bar\xi_{12} + \alpha \bar\xi_{22}
\eean
Consequently the periodicity properties of the various fields are
\bean
p_1&\rightarrow& p_1 -
\frac{\gamma}{\alpha\xi_{11}^2 + \gamma\xi_{11}\xi_{12}},\quad\quad\quad
\bar p_1 \rightarrow \bar p_1 -
\frac{\beta}{\delta{\bar\xi_{11}}^2 - \beta\bar\xi_{11}\bar\xi_{21}}\\
p_2&\rightarrow& p_2 +
\frac{\beta}{\delta\xi_{12}^2 + \beta\xi_{11}\xi_{12}},\quad\quad\quad
\bar p_2\rightarrow \bar p_2 +
\frac{\gamma}{-\gamma{\bar\xi_{11}}^2 + \alpha\bar\xi_{21}\bar\xi_{11}}
\eean
Therefore
for the pair $(p_1\, ,\,\bar p_2)$ we would have periodicity if
the condition $\gamma =0$ held true, while for the pair
$(p_2\, ,\,\bar p_1)$ we would have periodicity with $\beta =0$ and
finally for the pair $(p_2\, ,\,\bar p_2)$ to have periodicity the
condition $\gamma =\beta =0$ is required.
We will try to impose these conditions as dynamical constraints.
Therefore we have to calculate the Poisson brackets of $\beta$ and $\gamma$
with the various pairs above. We find
\begin{itemize}
\item
For the pair $(p_1\, ,\,\bar p_2)$
\[
\{p(x), \gamma\} = \gamma\delta(x)\, ,\;\;\;\;
\{\bar p(x) , \gamma\} = -\gamma\delta(x)
\]
\item For the second pair $(p_2\, ,\,\bar p_1)$
\[
\{p(x), \beta\} =-\beta\delta(x)\, , \;\;\;\;
\{\bar p(x) , \beta\} = \beta\delta(x)
\]
\item
For the last pair $(p_2\, ,\,\bar p_2)$
\a
\{p(x), \beta\} =-\beta\delta(x)\, ,
& &\{\bar p(x) ,\beta\} = \beta\delta(x)\0\\
\{p(x), \gamma\} = \gamma \delta(x)\, ,
& &\{\bar p(x) ,\gamma \} = -\gamma\delta(x)\0
\b
\item Moreover $\{\beta\, ,\,\gamma\} =0\, .$
\end{itemize}
The elements $\beta$  and $\gamma$ are preserved by taking the Poisson brackets
with the corresponding $p$'s, thus we can set them to zero as
hamiltonian constraints on the phase space. In this case the above
Poisson brackets tell us that the interesting physical quantities, i.e. the
fields $p$, are first class with respect to the constraints.

It remains for us to discuss periodicity.
Since in our setting the fields $p,\, {\bar p}$ are in principle not
periodic, we define the zero modes of the various fields to be
the integral over a period, namely we put
\[
p_{i,0}=\int_0^{2\pi}p_i (x)\,dx\, ,\;\;i=1,2
\]
and analogously, for ${\bar p}_i\,i=1,2.$ Using the linear systems
(\ref{Vconn}) and (\ref{barVconn}) we find
\bean
p_{1,0}&=&\log (\alpha +\gamma \frac{\xi_{21}(0)}{\xi_{11}(0)})\\
p_{2,0}&=&\log (\delta +\beta \frac{\xi_{11}(0)}{\xi_{12}(0)})\\
{\bar p}_{1,0}&=&-\log (\delta -
\beta\frac{\bar\xi_{21}(0)}{\bar\xi_{11}(0)})\\
{\bar p}_{2,0}&=&-\log (\alpha -
\gamma\frac{\bar\xi_{11}(0)}{\bar\xi_{21}(0)})
\eean
Imposing periodicity {\em and\/} equal zero
modes yields the following picture.
\begin{enumerate}
\item With $\gamma=0$ we have $p_1,\, {\bar p}_2$ periodic. Their zero
modes are equal, respectively, to $\log\alpha$ and $-\log\alpha .$
Setting them equal implies $\alpha ~=1,$ so that the monodromy must
be
\[
T=\left(\brr{cc}1&\beta\\0&1\err\right)
\]
\item with $\beta =0$ we have $(p_2,\, {\bar p}_1)$ to be the periodic
pair. The zero modes are $\log\delta$ and $-\log\delta ~,$ so that
$p_{2,0}={\bar p}_{1,0}$ implies
\[
T=\left(\brr{cc}1&0\\ \gamma &1\err\right)
\]
\item Setting simultaneously $\beta =0,\, \gamma=0$ all the fields
become periodic and the pair $p_{2,0},\, {\bar p}_{2,0}$ is
admissible. Equality of the zero modes translates into
$\log\delta =-\log\alpha$ which implies $\delta =1/\alpha ,$ so that
for the monodromy we have
\[
T=\left(\brr{cc}\alpha&0\\ 0&\frac{1}{\alpha}\err\right)
\]
\end{enumerate}

The last case, with diagonal real monodromy, is just
an unintersting subcase of the solutions discussed in section 3.
We disregard this case.

In the first two cases we have parabolic monodromy (and vanishing zero modes).
On the basis of the discussion
in section 3, all the solutions corresponding to the latter cases are
necessarily singular. Moreover they are not expected to be reconstructed
through a DS system of the type (\ref{DS+}, \ref{DS-}).
In conclusion we have obtained a parametrization of
such family of singular solutions in terms of free bosonic oscillators
(with a restricted phase space). We do not discuss here, for these solutions,
the locality property and the conjugate to the zero modes.

In Appendix D we present the same construction for the $sl_n$ case.

\subsection*{Appendix A. Hyperbolicity of the Monodromy: $sl_2$}

The monodromy of the Liouville theory ensuing from regular solutions (see below
for the precise meaning) is hyperbolic.
The proof goes as follows.
We exploit the representation of $T$ as a path-ordered exponential, but instead
of using the simplest contour,
namely the $t=0$ circle, we use the  following (closed) path
\[
\gamma_\tau=
\left\{ \brr{ll}(\tau ,\,\tau )&\tau\in [0,\pi ]\\
(\tau ,\, 2\pi -\tau )&\tau\in [\pi ,2\pi ]\err\right.
\]
which in the light-cone coordinates has the form
\[
\gamma_\tau=
\left\{ \brr{ll}(2\tau ,\, 0)&\tau\in [0,\pi ]\\
(2\pi ,\, 2\tau -2\pi )&\tau\in [\pi ,2\pi ]\err\right.
\]
Since this corresponds to a shift of $2\pi$ first in $x_+$ and then
in $x_-,$ we have
\[
T=\Psi_-\,\Psi_+
\]
with
\bean
\Psi_+&=&\mbox{{\bf P}}
\exp \left(-\int_0^{2\pi}A_+(x_+,x_-=0)\,dx_+\right)\\
\Psi_-&=&\mbox{{\bf P}}
\exp \left(-\int_0^{2\pi}A_+(x_+=2\pi ,x_-)\,dx_-\right)
\eean
Thus $\Psi_+$ is upper triangular, while $\Psi_-$ is lower triangular.
Finding the explicit form of $\Psi_\pm$ amounts to solving
\rref{LS} on the appropriate paths. The result is
\bean
\Psi_+ &=&
\left( \brr{cc}e^{-\frac{1}{2}(\varphi_1-\varphi_0)}
& -e^{-\frac{1}{2}(\varphi_0+\varphi_1)}
\int_0^{2\pi} e^{2\varphi (x_+,x_-=0)} dx_+\\
0& e^{\frac{1}{2}(\varphi_1-\varphi_0)}\err\right)
\\
\Psi_- &=&
\left( \brr{cc} e^{\frac{1}{2}(\varphi_2-\varphi_1)}&0\\
-e^{-\frac{1}{2}(\varphi_2+\varphi_1)}
\int_0^{2\pi}e^{2\varphi (x_+=2\pi ,x_-)}dx_-
&e^{-\frac{1}{2}(\varphi_2-\varphi_1)}
\err\right)
\eean
{}From this it is clear that in both cases the diagonal elements are
strictly positive while the off-diagonal ones are negative. We have
indicated with $\varphi_0,\,\varphi_1,\,\varphi_2$ the values taken by
$\varphi$ at the vertices of the triangle defined by the path. Notice
that periodicity implies that $\varphi_2=\varphi_0.$
For the trace we have
\bean
\mbox{tr}\,T&=&2\cosh\frac{1}{2}
(\varphi_2-2\varphi_1+\varphi_0)\,+\,\mbox{(positive contribution)}\\
&>&2
\eean
thus proving that in this case the monodromy $T$ is indeed hyperbolic. This
statement has been proved assuming that there are no singularities, but it is
easy to see that extending this method to more complicated zig--zag paths we
can
find a path that avoids isolated singularities and remains
homotopic to the initial one. However we must exclude singularities on
the $t=0$ axis and
accumulating singularities, continuous singular lines and all the like.
This accounts for the distinction between `regular' and `non regular'
solutions  throughout the article.

\subsection*{Appendix B. Hyperbolicity of the Monodromy: $sl_n$}

A matrix is hyperbolic when all its eigenvalues are strictly positive.
Here we prove that for regular solutions the monodromy
matrix for any $sl_n$ Toda field theory is hyperbolic.
We proceed as in the previous Appendix. We represent the monodromy matrix
in $sl(n)$ Toda theories
\a
T&=&P exp \left( \int_{0}^{2\pi} A_x(x,0) dx \right)\0
\b
as
\[
T=T_-(2\pi)T_+(2\pi)\0
\]
with
\bean
T_+(x_+)&=&\mbox{{\bf P}}
\exp \left(-\int_0^{x_+}A_+(\zeta _+,0)\,d\zeta_+\right)\0\\
T_-(x_-)&=&\mbox{{\bf P}}
\exp \left(-\int_0^{x_-}A_+(2\pi ,\zeta_-)\,dx_-\right)\0
\eean

Then we use the linear system (\ref{LS}) to obtain a system
of differential equations for the order $k$ ($k=1 \ldots n$) minors
of the matrices $T_{\pm}^{-1}(x_{\pm})$
\a
\partial_+ (T_+^{-1})_{\hskip 0.1 cm i_1 \ldots i_k}^{j_1 \ldots j_k}&=&
\left( \l_{j_1} +\ldots \l_{j_k}\right) (\partial_+ \Phi) \cdot
(T_+^{-1})_{\hskip 0.1 cm i_1 \ldots i_k}^{j_1 \ldots j_k}+\label{minorp}
\\
&&+e^{\al_{j_1-1}(\Phi)}   \cdot
(T_+^{-1})_{\hskip 0.1 cm i_1 \ldots i_k}^{j_1-1 \ldots j_k} +\ldots
+ e^{\al_{j_k-1}(\Phi)}   \cdot
(T_+^{-1})_{\hskip 0.1 cm i_1 \ldots i_k}^{j_1 \ldots j_k-1}\0
\b
\a
\partial_- (T_-^{-1})_{\hskip 0.1 cm i_1 \ldots i_k}^{j_1 \ldots j_k}&=&
-\left( \l_{j_1} +\ldots \l_{j_k}\right) (\partial_- \Phi) \cdot
(T_-^{-1})_{\hskip 0.1 cm i_1 \ldots i_k}^{j_1 \ldots j_k}+\label{minorm}\\
&&+e^{\al_{j_1}(\Phi)}   \cdot
(T_-^{-1})_{\hskip 0.1 cm i_1 \ldots i_k}^{j_1+1 \ldots j_k} +\ldots
+ e^{\al_{j_k}(\Phi)}   \cdot
(T_-^{-1})_{\hskip 0.1 cm i_1 \ldots i_k}^{j_1 \ldots j_k+1}\0
\b
where $\l_1 \ldots \l_n$ are the weights of the
defining representations of $sl(n)$.
Taking into account the initial conditions
\a
T_{\pm}(0)=1\0
\b
we obtain
\a
t_+^{ij}(x_+)&=&e^{\l_j(\Phi(x_+,0))-\l_i(\Phi(0,0))}\cdot \0\\
&&\hskip -1cm \cdot \int_0^{x_+}d\zeta_{j-1}
e^{2\al_{j-1}(\Phi(\zeta_{j-1},0))}
\int_0^{\zeta_{j-1}}d\zeta_{j-2} e^{2\al_{j-2}(\Phi(\zeta_{j-2},0))}
\ldots \int_0^{\zeta_{i+1}}
d\zeta_{i} e^{2\al_{i}(\Phi(\zeta_{i},0)}\0
\b
for $i<j$
\a
t_+^{ii}= e^{\l_j(\Phi(x_+,0))-\l_i(\Phi(0,0))}\0
\b
and
\a
t_+^{ij}=0\0
\b
for $i>j$. Moreover
\a
t_-^{ij}&=&e^{\l_i(\Phi(2\pi,0))-\l_j(\Phi(2\pi,x_-))}\cdot \0\\
&&\hskip -1cm  \cdot\int_0^{x_-}d\zeta_{j}
e^{2\al_{j}(\Phi(2\pi,\zeta_{j}))}
\int_0^{\zeta_j}d\zeta_{j+1}
e^{2\al_{j+1}(\Phi(2\pi,\zeta_{j+1}))} \ldots
\int_0^{\zeta_{i-2}}d\zeta_{i-1}
e^{2\al_{i-1}(\Phi(2\pi,\zeta_{i-1}))}\0
\b
for $i>j$
\a
t_-^{ii}&=&e^{\l_i(\Phi(2\pi,0))-\l_i(\Phi(2\pi,x)_-))}\0
\b
and
\a
t_-^{ij}=0\0
\b
for $i<j$.
Hence $t_{\pm}^{ij}\geq 0$.

In fact a much stronger result is
valid, namely, if $I=(i_1,i_2 \ldots i_k) \, \, \,
J=(j_1,j_2 \ldots j_k)$ are ordered multiindices
($ i_1<i_2< \ldots <i_k \, \, \, j_1<j_2 < \ldots j_k$), the
corresponding minors
\a
(T_{\pm}^{-1})^{IJ}=
(T_{\pm}^{-1})_{\hskip 0.1 cm i_1 \ldots i_k}^{j_1 \ldots j_k}
\label{multi}
\b
are nonnegative.
To show this we first note that
$(T_+^{-1})^{IJ}\neq 0$
only if $i_l\leq j_l$ ($l=1,2\ldots k$). A similar result is
also valid for the minors of $T_-^{-1}$,
$(T_-^{-1})^{IJ}\neq 0$
only if $i_l\geq j_l$ ($l=1,2\ldots k$). Moreover the following
factorization property holds
\a
(T_{\pm}^{-1})_{\hskip 0.1 cm i_1 \ldots i_k}^{j_1 \ldots j_k}=
(T_{\pm}^{-1})_{\hskip 0.1 cm i_1 \ldots i_{l-1}}^{j_1 \ldots j_{l-1}}
t_{\pm}^{i_lj_l}
(T_{\pm}^{-1})_{\hskip 0.1 cm i_{l+1} \ldots i_k}^{j_{l+1} \ldots j_k}
\quad\quad {\rm for}~~i_l=j_l\0
\b
The differential equations (\ref{minorp}) and (\ref{minorm})
and this identity allow us to express
$(T_{\pm}^{-1})^{IJ}$
as a sum of path ordered integrals all the integrands being
nonnegative functions. Therefore we conclude that (\ref{multi}) are
nonnegative.

It is also easy to verify that if $e^{\al_i(\Phi)}$
is not identically zero on the lines $0\leq x_+ \leq 2\pi , \, \,x_-=0 $
and  $x_+=2\pi, \, \, 0\leq x_- \leq 2\pi$, all the minors
$(T_{+}^{-1})^{IJ}$
( $(T_{-}^{-1})^{IJ}$)
are positive for $i_l\leq j_l$ ( $i_l\geq j_l$). Therefore
the minors
\a
(T^{-1})_{\hskip 0.1 cm i_1 \ldots i_k}^{j_1 \ldots j_k}&=&
\sum_{r_l\geq max(i_l,j_l)}
(T_{+}^{-1})_{\hskip 0.1 cm i_1 \ldots i_k}^{r_1 \ldots r_k}
(T_{-}^{-1})_{\hskip 0.1 cm r_1 \ldots r_k}^{j_1 \ldots j_k}\0\\
i_1<i_2<\ldots i_k && j_1<j_2<\ldots <j_k\0
\b
are positive.

To prove that all the eigenvalues of $T^{-1}$ are positive we introduce
the matrices $\T_k$ of order ${n\choose k} \times {n\choose k}$
with entries
\a
(\T_k)_{IJ}&=& (T^{-1})^{IJ}>0\0
\b
Let us denote by $\mu_1,\mu_2, \ldots \mu_n$ the eigenvalues of $T^{-1}$
ordered
as follows
\a
|\mu_1|\geq |\mu_2|\geq \ldots \geq |\mu_n|\0
\b
Hence the eigenvalues of $\T_k$ are
\a
\mu_{i_1},~ \mu_{i_2},~ \ldots \mu_{i_k} \0\\
i_1<i_2<\ldots <i_k\0
\b
The Perron-Frobenius theorem ( see \cite{Gant} vol. II, p.53 and 105)
states that a matrix with positive entries
has at least one positive eigenvalue which exceeds the moduli of all the
other eigenvalues. Applying this theorem to the matrices $\T_k$
we conclude that  $\mu_1 , \, \, \,  \mu_1 \mu_2 , \ldots,
\mu_1 \mu_2 \ldots \mu_{n}=detT^{-1}=1  $ are positive numbers
 and therefore all the eigenvalues of
$T^{-1}$ ( and so also of $T$) are {\it real positive numbers}.

The distinction made at the end of Appendix A between regular or non regular
solutions holds here as well.

\subsection*{Appendix C. Poisson algebra of $V$ and $\bar V$}.

The Poisson algebra of $V$ and $\bar V$ we need throughout the paper can
be calculated form (\ref{CPB}) with standard methods
\a
\{V(x) \x V(y) \} &=& \theta (x-y) V(x) \otimes V(y) \label{VV}\\
&&\cdot\left( r - V^{-1} (y) \otimes V^{-1}(y) \cdot (r-t_0)\cdot V(y)
\otimes V(y) \right)\0\\
&&+\theta(y-x) V(x) \otimes V(y)\0\\
&&\cdot \left( r - V^{-1} (x)\otimes V^{-1} (x)\cdot (r+t_0)\cdot  V(x)
\otimes V(x)\right) \0\\
\{V(x) \x \bar V(y) \} &=& \theta (x-y) V(x) \otimes 1\label{VVbar}\\
&&\cdot \left(- r + \bar V (y) \otimes \bar V(y) \cdot (r+t_0)\cdot
\bar V^{-1}(y) \otimes \bar V^{-1}(y) \right)1\otimes \bar V(y)\0\\
&&+\theta(y-x) V(x) \otimes 1\0\\
&&\cdot \left(- r + V^{-1} (x)\otimes V^{-1} (x)\cdot (r+t_0)\cdot  V(x)
\otimes V(x)\right)1\otimes \bar V(y) \0\\
\{\bar V(x) \x \bar V(y) \} &=&\left( r - \bar V (y) \otimes \bar V(y)
\cdot (r+t_0)\cdot \bar V^{-1}(y) \otimes \bar V^{-1}(y) \right)\0\\
&&\cdot\bar V(x) \otimes \bar V(y)  \theta (x-y) \label{VbarVbar}\\
&& + \left( r - \bar V (x) \otimes \bar V(x) \cdot (r-t_0)\cdot \bar V^{-1}(x)
\otimes \bar V^{-1}(x) \right)\0\\
&&\cdot \bar V(x) \otimes \bar V(y)\theta (y-x)\0
\b

\subsection*{Appendix D. Explicit construction of the DS fields: $sl_n$}

In this Appendix we do for $sl_n$ the same construction as in section 5
where we constructed the $p$ and $\bar p$ fields corresponding to
singular solutions of the Liouville equations.

After the gauge transformation
\a
V&=&G_1G_2 \ldots G_{n-1} \V_+ \0\\
G_i&=&e^{X_{i+1 i}E_{i+1 i}} e^{X_{i+2 i}E_{i+2 i}} \ldots e^{X_{ni}E_{ni}}
\b
the transformed transport matrix $\V_+$ satisfies DS type
linear system (\ref{DS+}) iff $X_{ij}\,, \, i>j$,
satisfy the equations:
\a
\partial_+ X_{ij}&=&2X_{ij}\sum_{k=j}^{i-1}\alpha_k(\partial_+ \Phi)
-X_{ij}X_{j j-1} +X_{i j}X_{j+1 j}+X_{i j-1}-X_{i+1 j}\0\\
\partial_-X_{ij}&=& -e^{2\alpha_{i-1}(\Phi)} X_{i-1j} \0\\
 X_{ii}&=&1
\label{gauge}
\b
The Toda equations appear both as integrability conditions of
(\ref{gauge}) and in the chirality of the fields
\a
P= -2\partial_+ \Phi -
\sum_{i=1}^{n-1}X_{i+1 i}H_{\alpha_i}
\b
Similarly in order to get  the antichiral fields
we consider the gauge transformation
\a
\bar{V}&=&\V_-\bar{G}_{n-1} \bar{G}_{n-2}
\ldots \bar{G}_1 \0\\
\bar{G_i}&=&e^{\bar{X}_{ii+1}E_{ii+1}} e^{\bar{X}_{ii+2}E_{ii+2}}
\ldots e^{\bar{X}_{in}E_{in}} \0
\b
and the new transport matrix
$\V_-$ satisfy (\ref{DS-}) iff $\bar{X}_{ij \, ,\, i<j}$,
satisfy the following system of differential
equations
\a
\partial_+ \bar{X}_{ij}&=&e^{2\alpha_{j-1}(\Phi)}X_{i j-1} \0\\
\partial_- \bar{X}_{ij}&=&2\bar{X}_{ij}
\sum_{k=i}^{j-1}\alpha_k (\partial_- \Phi)+
\bar{X}_{i-1i} \bar{X}_{ij}-\bar{X}_{ii+1}\bar{X}_{ij}-
\bar{X}_{i-1j}+\bar{X}_{ij+1} \0\\
\bar{X}_{ii}&=&0
\label{bargauge}
\b
The Toda equations again appear both as integrability conditions of
(\ref{bargauge}) and in the chirality of the fields
\a
\bar{P}&=& 2\partial_- \Phi -\sum_{i=1}^{n-1}\bar{X}_{ii+1}H_{\alpha_i}
\b
Let us now see how this construction works in the example of
the defining representation of $sl(n)$. Denote by $\xi_{ij}$
($\bar{\xi}_{ij}$) the matrix elements of the transport matrices
$V$ ($\bar{V}$) and introduce the notation
\be
V_{\hskip 0.1 cm i_1 \ldots i_k}^{j_1 \ldots j_k}\,=\,
det \left(\brr{ccc}\xi_{i_1j_1}&\ldots & \xi_{i_1 j_k} \\
           \ldots &  \ldots & \ldots  \\
\xi_{i_kj_1} & \ldots &\xi_{i_kj_1} \err\right) ,\quad\quad
\bar{V}_{\hskip 0.1 cm i_1 \ldots i_k}^{j_1 \ldots j_k}\,=\,
det \left(\brr{ccc}\bar{\xi}_{i_1j_1}&\ldots & \bar{\xi}_{i_1 j_k} \\
           \ldots &  \ldots & \ldots  \\
{\xi}_{i_kj_1} & \ldots &{\xi}_{i_kj_1} \err\right)
\ee
For arbitrary permutations\footnote{We use the notation $\pi(i)=\pi_i$, etc.}
$\pi$ and $\sigma$ the matrices
$\V_+^{\pi} $ and $\V_-^{\sigma}$ with entries
\a
\hat{\xi}_{ij}^{\pi}\,= \, \textstyle
{V^{\pi_ 1 \ldots \pi_{i-1} \pi_i}_{1\ldots i-1 j}  \over
V^{\pi_1\ldots  \pi_{i-1}}_{1\ldots i-1} } ,\quad\quad
\hat{\bar{\xi}}_{ij}^{\sigma}\,= \,
\textstyle{ \bar{V}_{\sigma_1\ldots \sigma_{j-1}
 \sigma_i}^{1\ldots \hskip 0.1 cm j-1 \hskip 0.15 cm j}
\over \bar{V}_{\sigma_1\ldots \sigma_{j-1}}^{1\ldots \hskip 0.1 cm  j-1}}
\label{hatKsiij}
\b
satisfy the linear systems (\ref{DS+}) and (\ref{DS-}). The
chiral (antichiral) momenta $P^{\pi}=diag(p_1^{\pi},\ldots , p_n^{\pi})$
$( \bar{P}^{\sigma}=diag(\bar{p}_1^{\sigma},\ldots , \bar{p}_n^{\sigma})$
are given by
\a
p_i(x)^{\pi}=\partial_x \log \hat{\xi}_{ii}^{\pi}  ,\quad\quad
\bar{p}_i^{\sigma}(x) -\partial_x log  \hat{\bar{\xi}}_{ii}^{\sigma}
\b

Then using the Poisson brackets of the theory (see Appendix C) one obtains
\a
\{ p_k^{\pi}(x) , p_l^{\pi}(y) \}=
- \{ \bar{p}_k^{\pi}(x) , \bar{p}_l^{\pi}(y) \} =
=\partial_x \partial_y \epsilon (x-y) \sum_{j=1}^{n-1}
\l_{\pi_k} (H_j) \l_{\pi_l}(H_j)
\b
On the other hand we have to impose also
\a
\{ p_k^{\pi}(x) , \bar{p}_l^{\sigma}(y) \}=0
\b
This has three classes of solutions
\begin{itemize}
\item
The pairs $(\pi\, ,\,\sigma )$ with
\[
\pi_i=\sigma_{n-i+1}
\]
\item The pairs $(\pi\, ,\,\sigma)$ with
\[
\pi_i=n-i+1
\]
and $\sigma$ arbitrary.
\item
The pairs $(\pi\, ,\,\sigma)$
\[
\sigma_i=n-i+1
\]
and $\pi$ arbitrary.

\end{itemize}

One can prove that, among these solutions, there exist those for which
we can impose, via hamiltonian constraints, that the monodromy matrix
be upper triangular (or lower triangular) and the diagonal elements
be such that be non diagonalizable. These solutions are necessarily
singular. For them it is possible to define a free field representation.


\begin{thebibliography}{}

\bibitem{GN} J.-L.Gervais and A.Neveu, Nucl.Phys. B199 (1982) 59; B209 (1982)
125; B238 (1984) 125, 396.
\bibitem{CT} T.L.Curtright and C.B.Thorn, Phys.Rev.Lett. 48 (1982) 1309;
E.Braaten, T.L.Curtright and C.B.Thorn, Phys.Lett. 118B (1982) 115;
Ann.Phys. 147 (1983) 365.
\bibitem{S} N.Seiberg, {\it Notes on quantum Liouville theory and quantum
gravity}, lecture notes RU--90--29; L.Alvarez--Gaum\'e, Helv.Phys.Acta 64
(1991) 361; and references therein.
\bibitem{FL} V.A.Fateev and S.L.Lukyanov, Int.Jour.Mod.Phys. A3 (1988),
507, A7 (1992), 853.
\bibitem{GB} J.-L.Gervais and A.Bilal, Nucl.Phys. B314 (1989) 646,
               B318 (1989) 579.
\bibitem{B} O.Babelon, Phys.Lett. B215 (1988) 523.
\bibitem{GM} J.-L. Gervais and Y.Matsuo, LPTENS--91/35.
\bibitem{BBT} O.Babelon, L.Bonora and F.Toppan, Comm.Math.Phys. 140 (1991) 93.
\bibitem{LS} A.N.Leznov and M.V.Saveliev, Lett.Math.Phys. 3 (1979), 489.
\bibitem{AB} E.Aldrovandi and L.Bonora, in preparation.
\bibitem{DS} V.G.Drinfeld and V.V.Sokolov, J.Sov.Math. 30 (1984) 1975.
\bibitem{FT} L.D.Faddeev and L.Takhtajan, Springer Lecture Notes in Physics,
vol. 246 (1986) 166.
\bibitem{Gant} F.R.Gantmacher, {\it The theory of matrices}, New York,
1969.

\end{thebibliography}
\end{document}